\documentclass{osa-article}

\journal{oe}

\articletype{Research Article}

\usepackage{lineno}

\begin{document}

\title{Development of hard masks for reactive ion beam angled etching of diamond}

\author{Cleaven Chia,\authormark{1,2,\textdagger} Bartholomeus Machielse,\authormark{3,4,\#} Amirhassan Shams-Ansari,\authormark{1} and Marko Lon\v{c}ar\authormark{1,*}}

\address{
\authormark{1}John A.\ Paulson School of Engineering and Applied Sciences, Harvard University, Cambridge, Massachusetts 02138, USA\\
\authormark{2}Current affiliation: Institute of Materials Research and Engineering (IMRE), A*STAR (Agency for Science, Technology and
Research) Research Entities, 2 Fusionopolis Way, \#08-03 Innovis, 138634, Singapore\\
\authormark{3}AWS Center for Quantum Computing, Pasadena, CA 91125, USA \\
\authormark{4}Department of Physics, Harvard University, Cambridge, Massachusetts 02138, USA }

\email{\authormark{*}loncar@seas.harvard.edu} 



\begin{abstract}
Diamond offers good optical properties and hosts bright color centers with long spin coherence times. Recent advances in angled-etching of diamond, specifically with reactive ion beam angled etching (RIBAE), have led to successful demonstration of quantum photonic devices operating at visible wavelengths. However, larger devices operating at telecommunication wavelengths have been difficult to fabricate due to the increased mask erosion, arising from the increased size of devices requiring longer etch times. We evaluated different mask materials for RIBAE of diamond photonic crystal nanobeams and waveguides, and how their thickness, selectivity, aspect ratio and sidewall smoothness affected the resultant etch profiles and optical performance. We found that a thick hydrogen silesquioxane (HSQ) layer on a thin alumina adhesion layer provided the best etch profile and optical performance. The techniques explored in this work can also be adapted to other bulk materials that are not available heteroepitaxially or as thin films-on-insulator.
\end{abstract}

\section{Introduction}
Diamond is a promising material for realizing photonic devices due to its high refractive index (n = 2.4), its wide bandgap (5.5 eV) and transparency window, and large-power handling capability. Leveraging these properties, devices such as nanowires, \cite{BabinecNNano2010} gratings, \cite{ForsbergDRM2013,MakitaMicroelectronEng2017,KissOE2019} mirrors, \cite{AtikianarXiv2019} and metasurface frequency converters \cite{Shen2021} have been realized. Diamond also has strong third order optical nonlinearity which has been leveraged to realize frequency combs, \cite{HausmannNPhot2014} Raman lasers, \cite{LatawiecOptica2015,LatawiecOL2018} and supercontinuum sources. \cite{Shams-AnsariOL2019} Perhaps the most exciting application of diamond is in quantum photonics enabled by a wide variety of color centers that exist within diamond’s wide bandgap, featuring bright coherent optical transitions and long spin coherence times. A good example is the negatively-charged silicon-vacancy center (SiV\textsuperscript{--}) that has been integrated into quantum photonic devices such as photonic crystal nanobeams. \cite{EvansScience2018} These were subsequently used to demonstrate coupling between SiV\textsuperscript{--} centers and photons with high cooperativity, \cite{NguyenPRL2019} paving the way for the creation of quantum memory nodes \cite{NguyenPRB2019} and memory-enhanced quantum communication. \cite{BhaskarNature2020} To produce high-performance quantum photonic devices in diamond, tight confinement of optical modes with low optical loss is needed. These requirements can be met if diamond devices are supported by lower-index substrates or are suspended in air (free-standing). 

One often pursued approach relies on fabrication of thin diamond films -- free-standing or on low index substrates. In this technique, devices are etched into thin films, which can be created either by using a combination of laser slicing and reactive ion etching, or using ion implantation and electrochemical etching. Recently, a variant of this approach has been developed that uses diamond regrowth and fusion bonding before electrochemical etching to remove the damaged layer. \cite{WangJVSTB2007,PirachaNL2016,kuruma2021telecommunication}

\begin{figure}
\centering\includegraphics[width=\textwidth,keepaspectratio]{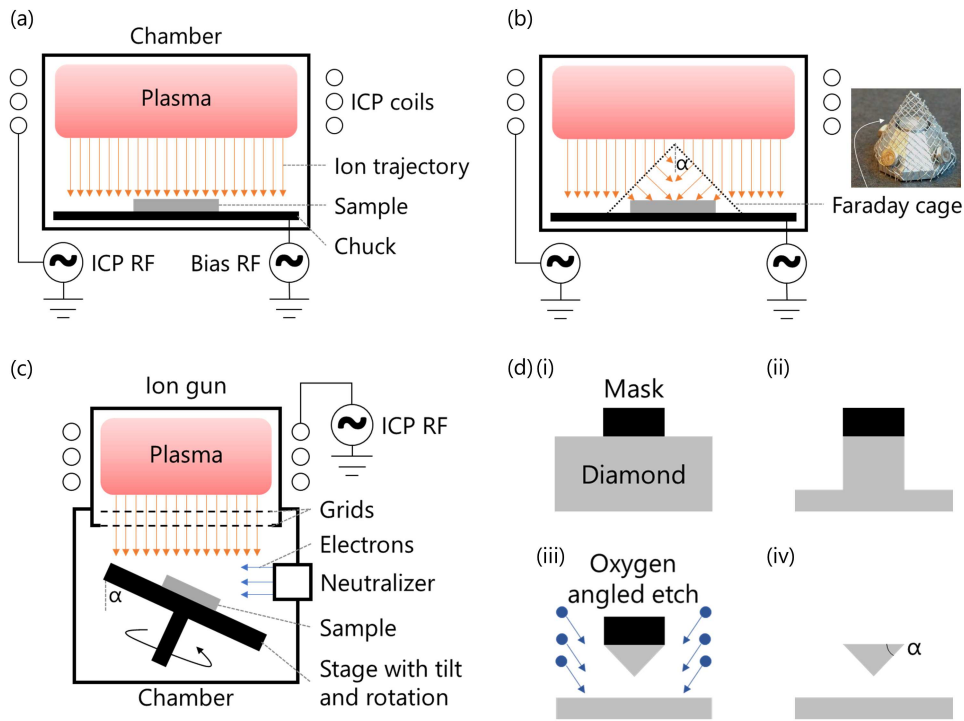}
\caption[Comparison between conventional reactive ion etching (RIE), angled RIE, and ion beam etching, with process flow for angled etching]{Comparison between conventional reactive ion etching (RIE), angled RIE, and ion beam etching. (a) Schematic for conventional reactive ion etching (RIE). (b) Schematic of angled RIE using a Faraday cage to modify ion trajectories after entering the cage. \(\alpha\) is the half-apex angle of the cage. Right: Photograph of Faraday cage with sample mounted inside it. (c) Schematic of ion beam etching, with stage tilted at angle \(\alpha\) relative to ion beam. (d) process flow for angled etching: (i) mask definition, (ii) anisotropic etching, (iii) angled etching, (iv) mask removal. \(\alpha\) defines the etch angle, related to the Faraday cage half-apex angle in (b) and the stage angle in (c).
\label{fig:FigFab1-RIE_FC_IBE_process}}
\end{figure}

An alternative approach relies on machining of bulk diamond substrates using either angled etching or quasi-isotropic etching to realize free-standing diamond devices. Despite great promise and versatility offered by isotropic undercut approach, \cite{KhanalilooNL2015,MitchellOptica2016,MouradianAPL2017,WanAPL2018,DoryNComm2019} including recent work on device transfer and large scale integration, \cite{WanNature2020} state-of-the-art-quantum photonic devices are still made using angled-etching approach. \cite{NguyenPRB2019,NguyenPRL2019,BhaskarNature2020} The first angled etching of diamond was done by housing the diamond substrate in a home-made Faraday cage within an inductively-coupled plasma reactive ion etching (ICP-RIE) chamber. \cite{BurekNL2012,BurekNComm2014,LatawiecJVSTB2016} The Faraday cage modifies the trajectories of ions in the chamber: instead of travelling at normal incidence to the sample [Fig. \ref{fig:FigFab1-RIE_FC_IBE_process}(a)], they are redirected normal to the Faraday cage boundaries, and hence they travel at oblique incidence to the sample [Fig. \ref{fig:FigFab1-RIE_FC_IBE_process}(b)]. This results in undercutting of diamond nanostructures and produces devices with triangular cross sections. Since then, Faraday cage angled etching has also been used to etch nanocones \cite{JeonACSPhot2020} and multi-layered photonic crystal mirrors \cite{JeonOE2020} in diamond, as well as nanobeams in other materials such as quartz, \cite{SohnAPL2016} gallium nitride, \cite{GoughAIPAdv2020} and silicon carbide. \cite{CranwellAdvMatTech2021,babin2022fabrication} However, the Faraday cage does not redirect the ion trajectories uniformly, leading to varying degrees of undercut across the substrate. \cite{LatawiecJVSTB2016} As a result, devices produced by Faraday cage angled etching are not uniform across the substrate, limiting the reproducibility on a single substrate and from run to run.

More recently, angled etching of diamond devices has been performed via reactive ion beam angled etching (RIBAE). \cite{AtikianAPLPhot2017} In this method, the ion beam is generated in an ion gun external to the etching chamber, and collimated and accelerated towards the chamber through a series of grids [Fig. \ref{fig:FigFab1-RIE_FC_IBE_process}(b)]. Importantly, the ion beam can be made uniform over an area of diameter as large as 4-inch, in the case of the tool used in this work. This, combined with automatic stage tilt and continuous rotation during the etch process, ensures improved undercut etch uniformity over the entire area of a bulk diamond substrate, which is on the order of 4 mm \(\times\) 4 mm. RIBAE has been used successfully to fabricate photonic devices like diamond mirrors, \cite{AtikianarXiv2019} dispersion engineered waveguides for supercontinuum generation, \cite{Shams-AnsariOL2019} and photonic crystal cavities used in cavity QED experiments. \cite{NguyenPRB2019,NguyenPRL2019,BhaskarNature2020} The main challenge of the angled etching approach is mask erosion and deformation, more significant than in vertical etches, that ultimately limits the width of the devices that can be realized. This is because in angled etching, the incident ion beam impinges on the mask at an oblique angle, and the sidewalls of the mask sees a non-zero ion flux. As a result, mask material is sputtered from its sides, leading to lateral mask erosion. In addition, for amorphous or polycrystalline materials typically used for masks, sputtering yields at oblique incidence is higher than at normal incidence \cite{Wei_2008}, leading to greater vertical mask erosion rates in angled etching compared to conventional top-down etching. Furthermore, mask erosion ultimately impacts the Q of fabricated devices by modifying cross section shapes. Consequently, while devices operating in visible could readily be achieved, larger devices operating at telecommunication wavelengths have been elusive given the greater extent of mask erosion arising from longer etch times. 

To address these challenges, we compared different mask materials suitable for RIBAE. We examined the requirements for different nanophotonic geometries, such as photonic crystal nanobeam cavities and waveguides. These results provide insights into the nature of lateral erosion of masks during angled diamond etching, enabling improved optical performance of fabricated devices. We also benchmark the measured optical Q of devices fabricated in RIBAE with different mask materials against previous devices fabricated using Faraday cage angled etching. \cite{BurekOptica2016} Furthermore, the results of this study can be adapted to any material for which no thin film platform is available, and where fabrication must be done from bulk substrates.

We begin by offering a general analysis of the problem of angled ion beam etching of photonic devices. A proper mask material for angled etching processes should be sufficiently thick, have high selectivity, good aspect ratio tolerance, high resolution and smooth sidewalls. A sufficiently thick mask is required to withstand erosion during both vertical and angled etches. The mask should also have a high selectivity to diamond during the etching processes, since low selectivities lead to high vertical and lateral erosion rates in both vertical and angled etches, which will change the mask shape and deform the final device geometry. For photonic crystal nanobeams, which are nanobeam waveguides with quasi-periodic perforations along its length, the mask needs to be made thick enough to protect the incident ion beam from etching into the perforations and inducing sidewall roughness. As these geometries have features with small lateral dimensions relative to the thickness, the mask would need a good tolerance to high aspect ratios; this also requires high resolution in the lithography process for the mask to be accurately defined throughout the depth of the mask. Finally, smooth sidewalls are needed for the mask material, as roughness in the mask that is transferred into diamond during the etching processes will result in higher optical scattering loss rates.

We evaluated several combinations of mask materials in this work: (i) hydrogen silesquioxane (HSQ) on titanium (Ti) adhesion layer (hereafter shortened to HSQ-Ti); (ii) HSQ on niobium (Nb) adhesion layer, (HSQ-Nb); (iii) templated Nb or alumina using poly(methyl methacrylate) (PMMA); and (iv) HSQ on alumina adhesion layer (HSQ-alumina).

\section{Device design, fabrication, and measurement} \label{chap:ibe-sec:fabprocess}
The fabrication process we used to evaluate different mask materials consists of four steps [Fig. \ref{fig:FigFab1-RIE_FC_IBE_process}(d)]: defining the mask material, vertical etching to transfer the mask pattern into diamond, angled etching to undercut diamond to suspend devices above the substrate, and removing the mask material to produce the final suspended devices. In a device with triangular cross section, its thickness \(t\) is related to its width \(w\) through the etch angle \(\theta\), via the relation \(t = w/(2  tan  \theta)\). To fully suspend the structures, the vertical etch depth should theoretically be at least the final thickness of the device. In practice, the etch depth is chosen to be larger to take into account evanescent leakage of light into the substrate for nanophotonic devices, and non-uniformity of etching photonic crystal devices in particular, where vertical etch rates in narrow features are lower than that in the wider ones.

Prior to mask definition, the diamond substrate was cleaned in 49\% hydrofluoric acid (HF) for 5 minutes, piranha solution (96\% sulfuric acid and 30\% hydrogen peroxide in 3:1 ratio) for 5 minutes, then ultrasonic agitation in acetone followed by methanol, and finally dried with a nitrogen gun. Following the mask definition process, which will be explained in further details for each mask material, a vertical oxygen etch was performed in a PlasmaTherm Versaline ICP-RIE system to transfer the mask pattern into diamond. This recipe used a flow of 40 standard cubic centimeters per minute (sccm), pressure of 10 milliTorr, bias and ICP powers of 100 W and 700 W. 

Angled etching with RIBAE was done with an Intlvac Nanoquest Ion Beam Etching System using oxygen ion beams from a Kaufman \& Robinson ion beam source. Two recipes were used in this work: (i) beam voltage 200 V, accelerator voltage 26 V, beam current 100 mA, ICP power 170 W, oxygen flow 38 sccm; (ii) beam voltage 150 V, accelerator voltage 22 V, beam current 80 mA, ICP power 130 W, oxygen flow 30 sccm. For both recipes, the stage angle was \(\alpha\) = 45\({}^{\circ}\).

For mask removal, the diamond substrate with devices was immersed in a solution of 49\% HF and 60\% nitric acid in 1:1 ratio for 5 minutes, followed by piranha solution for 5 minutes, and lastly dried through a critical point drying process with carbon dioxide.

The photonic crystal devices in this work consist of a clamp on one end, followed by the photonic crystal, a tapered support to provide tethering to the substrate via a thin fin that does not induce coupling losses, \cite{BurekNComm2014} and finally a tapered waveguide for adiabatic fiber coupling to tapered fibers fabricated by pulling out of HF solution. \cite{BurekPRAppl2017,TieckeOptica2015,Turner1984} These tapered fibers were placed in contact with the tapered diamond waveguide, using motorized stages, for efficient adiabatic coupling of light into the photonic crystal. The photonic crystal is designed to support a fundamental quasi-transverse electric (TE) mode near 1550 nm, and has a less reflective mirror on the end closer to the tapered waveguide than the end closer to the clamp, so as to increase coupling through the waveguide. Optical reflection spectroscopy was performed by coupling light from a tunable laser source (Santec TSL) and measuring the resultant reflection from the photonic crystal using a photodetector (New Focus). 

Racetrack resonators in this work are designed to have resonances in the telecommunication wavelength range, and are tethered to the substrate by a thin fin at the bottom apex of the cross section. This provides structural stability while minimizing optical losses into the substrate. \cite{BurekNComm2014} Optical Q’s of fabricated ring resonators were measured by coupling light from a tunable laser source (Santec TSL) evanescently through a tapered fiber (ThorLabs), \cite{BurekNComm2014} which was tapered to a diameter of \(\sim\) 1 \(\mu\)m by pulling under a hydrogen flame. The position of the tapered fiber was controlled with motorized stages and optimized for critical coupling to the ring resonator. The resulting transmission was measured with a photodetector (New Focus). \cite{AtikianAPLPhot2017,BurekNComm2014}

\section{Results for different mask methods}
\subsection{HSQ-Ti mask}
\begin{figure}
\centering\includegraphics[width=\textwidth,keepaspectratio]{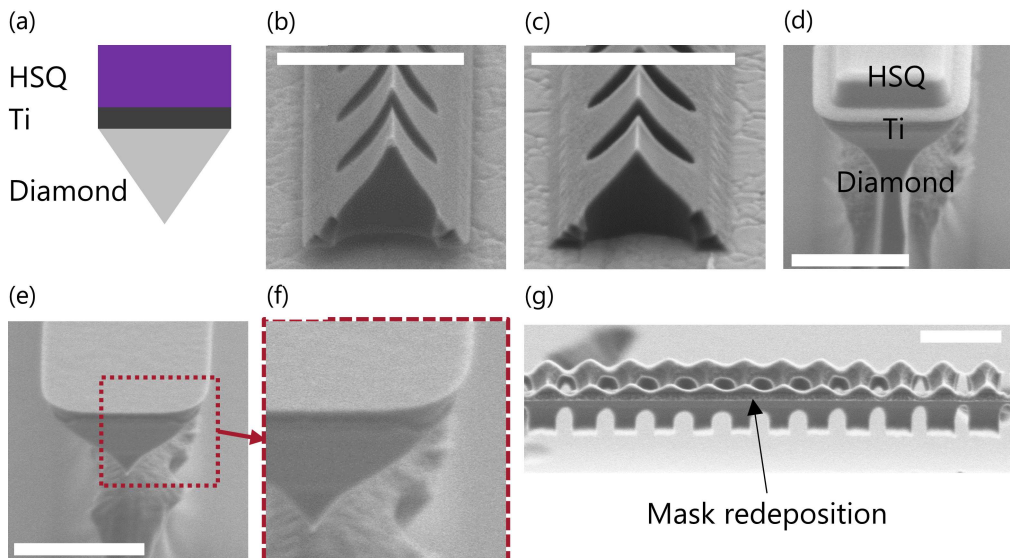}
\caption[Angled etching using HSQ-Ti mask]{Angled etching using hydrogen silesquioxane (HSQ)-Ti mask. (a) Schematic of HSQ-Ti mask stack for angled etching. (b) Scanning electron microscope (SEM) image of asymmetric cross-section from Faraday cage angled etching without cage rotation. (c) SEM image of a symmetric cross-section with Faraday cage rotation between etches. (d) SEM image of beam cross-section during RIBAE with continuous stage rotation. (e) SEM image of symmetric beam cross-section after RIBAE with continuous stage rotation. (f) Close up view of the cross-section in (e), showing mask redeposition near the top edge. (g) SEM image of photonic crystal after angled etching with HSQ-Ti mask, showing excessive mask erosion and redeposition. Scale bars in (b) – (e), (g) represent 1 \(\mu\)m. All SEM images were taken at a 60\({}^{\circ}\) tilt.
\label{fig:FigFab2-HSQTi}}
\end{figure}

We first considered a mask with hydrogen silesquioxane (HSQ; FOx-16, from Dow-Corning) and titanium (Ti) [Fig. \ref{fig:FigFab2-HSQTi}(a)]. HSQ is a negative-tone electron beam resist that cures to form silicon dioxide after electron beam lithography (EBL) and development, while Ti serves as an adhesion layer between HSQ and diamond, as well as a charge compensation layer for the lithography process. After cleaning of the diamond substrate, the mask definition proceeded first through electron beam evaporation (Denton) of a 40 nm layer of Ti, followed by spin-coating of HSQ. For fabricating photonic crystals operating in the telecommunication wavelength range (around 1550 nm), a 1 \(\mu\)m layer of HSQ was spin-coated to shadow the holes from being etched during the angled etching process.
This was followed by electron beam lithography (125 keV, Elionix) and development in 25\% tetramethylammonium hydroxide (TMAH) solution. Prior to the vertical oxygen etch, the Ti layer was etched in the aforementioned PlasmaTherm Versaline ICP-RIE with argon/chlorine (Ar/Cl\({}_{2}\)) for 1 minute with gas flows of 25 and 40 sccm respectively, pressure of 8 milliTorr, bias and ICP powers of 250 W and 400 W respectively.

The automatic stage rotation during RIBAE improved the etch cross-section uniformity. Previously, with Faraday cage angled-etching, a lack of stage rotation during the etch would result in asymmetric cross-sections [Fig. \ref{fig:FigFab2-HSQTi}(b)]. This was only improved by performing multiple etches of shorter durations, with the stage rotated manually after each etch was complete and the sample was unloaded [Fig. \ref{fig:FigFab2-HSQTi}(c)]. With automatic stage rotation during RIBAE, the cross-section was guaranteed to be symmetric [Fig. \ref{fig:FigFab2-HSQTi}(e)]. While the HSQ-Ti mask stack has advantages in terms of smoothness, resolution and thickness, it suffered from high top-down and lateral erosion rates [Fig. \ref{fig:FigFab2-HSQTi}(d)], as well as excessive mask redeposition [Fig. \ref{fig:FigFab2-HSQTi}(f)].

With RIBAE using the recipe with 200 V beam voltage, the HSQ-Ti mask stack has been used to angle-etch racetrack resonators with optical Q’s of 286 000, \cite{AtikianAPLPhot2017} as well as photonic crystals in visible wavelengths. \cite{NguyenPRB2019,EvansScience2018,MachielsePRX2019} For waveguides, the mask needs to be sufficiently thick and wide to withstand lateral and top-down erosion of the HSQ mask. For photonic crystals, there is an added requirement for the mask to be thick enough to shadow the holes from being etched. However, the maximum realizable aspect ratio of HSQ limits the range of thickness where features can be reliably reproduced.

While visible-wavelength photonic crystals have been successfully fabricated using HSQ-Ti mask stack and used to demonstrate memory-enhanced quantum communication, \cite{NguyenPRB2019,NguyenPRL2019,BhaskarNature2020} the same mask stack was unsuccessful in producing photonic crystals at telecommunication wavelengths. Firstly, limitations in aspect ratios that can be achieved with HSQ imposes an upper limit on its thickness, above which features cannot be faithfully reproduced with lithography. Secondly, the larger widths of telecommunication-wavelength photonic crystals as compared to their visible-wavelength counterparts means that a longer duration of RIBAE is needed to fully suspend the photonic crystals. This leads to a greater extent of mask erosion as well as mask redeposition, contributing to roughness in both outer edges of waveguides and inner sidewalls of holes [Fig. \ref{fig:FigFab2-HSQTi}(g)].

\subsection{HSQ-Nb mask} \label{chap:ibe-subsec:HSQ-Nb}
To overcome the large extent of mask erosion arising from HSQ mask, we explored the use of a metal mask for oxygen RIBAE, as many metals form a resistant oxide layer upon oxidation. Metals such as chromium, \cite{HodgesNJP2012} aluminum, \cite{ForsbergDRM2013,HwangDRM2004,HicksSciRep2019} and niobium  \cite{AtikianAPLPhot2017,AtikianarXiv2019} have previously been used as metal masks for diamond etching. Here, we chose niobium (Nb) due to its high selectivity and low erosion rate, as well as its ease of etching in Ar/Cl\({}_{2}\) etch.

\begin{figure}
\centering\includegraphics[width=\textwidth,keepaspectratio]{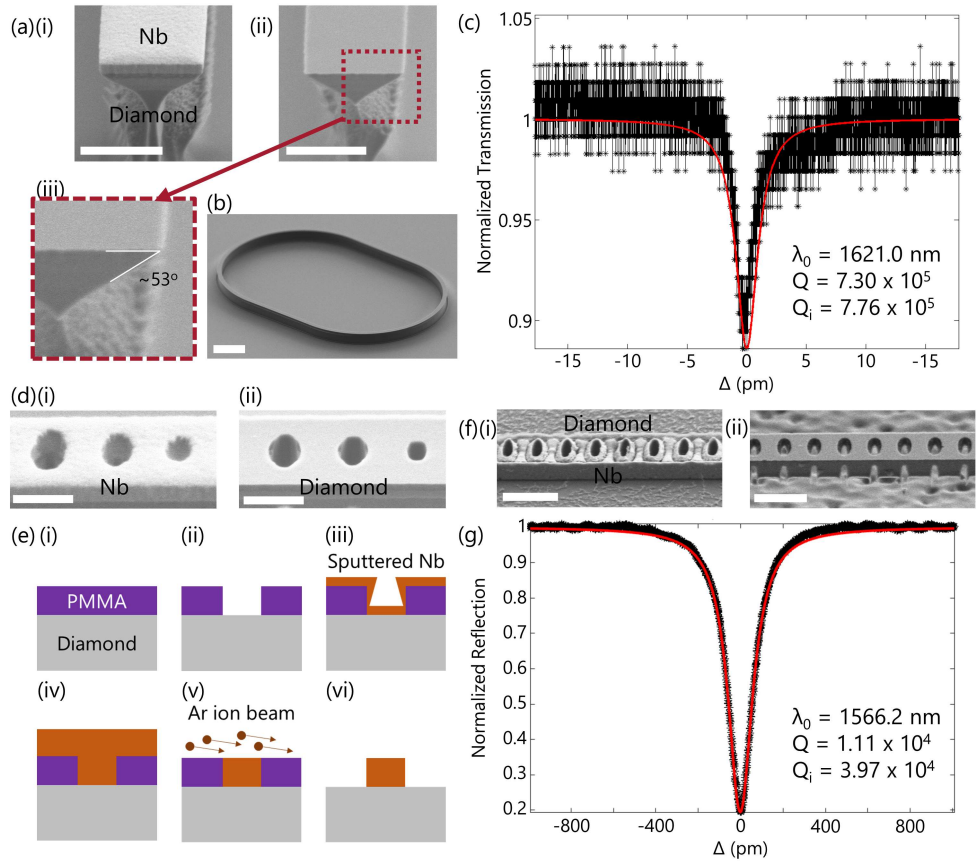}
\caption[RIBAE of photonic structures using Nb mask]{RIBAE of photonic structures using Nb mask. (a)(i) Scanning electron microscope (SEM) image during RIBAE of the cantilever with HSQ-Nb mask. (ii) SEM image of a completely etched cantilever. (iii) Close-up view of the cross-section of the cantilever, with a final etch angle close to 53\({}^{\circ}\). (b) SEM image of racetrack resonator fabricated with HSQ-Nb mask. (c) The optical transmission spectrum of racetrack resonator shown in (b), with optical resonance at 1621.0 nm, total optical Q of 730 000, and intrinsic optical Q of 776 000. (d)(i) SEM image of HSQ-Nb mask of photonic crystal holes, with HSQ removed to reveal grain structure in Nb mask. (ii) SEM image of rough hole sidewalls in diamond photonic crystal etched with HSQ-Nb mask. (e) Process flow for Nb mask on diamond via inversion of PMMA template (referred to as PMMA/Nb). (i) Spin coating of PMMA, (ii) electron beam lithography and development of PMMA template, (iii) sputtering of Nb into PMMA template, (iv) completed sputtering with Nb fully planarized over PMMA template, (v) planarization of Nb with argon ion beam at grazing incidence angle, (vi) final Nb mask on diamond. (f)(i) SEM image after removal of PMMA revealed Nb mask. (ii) SEM image of photonic crystal after angled etching with PMMA/Nb mask. (g) The optical reflection spectrum of photonic crystal fabricated with PMMA/Nb mask, with optical resonance at 1566.2 nm, total optical Q of 11 100, and intrinsic optical Q of 39 700. Scale bars represent 1 \(\mu\)m in (a) and (f), 10 \(\mu\)m in (b), and 500 nm in (d). All SEM images were taken at a 60\({}^{\circ}\) tilt.
\label{fig:FigFab4-PMMANb}}
\end{figure}

To fabricate diamond photonic devices with Nb masks, a \(\sim\) 200 nm layer of Nb was sputtered using a magnetron sputtering system with a 2-inch niobium target (AJA International) onto cleaned diamond substrates, and a 1 \(\mu\)m layer of HSQ was then spin-coated on top of niobium. Following EBL and development, the mask pattern defined in HSQ is transferred into the niobium layer with an anisotropic Ar/Cl\({}_{2}\) etch, and then transferred into the diamond substrate using an anisotropic oxygen etch for 10 mins. This was then followed by oxygen RIBAE with 200 V beam voltage to suspend the photonic structures, and mask removal to remove HSQ and niobium. The lateral etch rate of diamond during RIBAE was approximately 3.8 nm/min. Although HSQ was not removed prior to angled etching, it did not serve as the primary mask material here as it was eroded at a faster rate compared to niobium during the oxygen RIBAE step [Fig. \ref{fig:FigFab4-PMMANb}(a)(i)].

We first used the HSQ-Nb mask to etch cantilevers, using the recipe with 200 V beam voltage, and observed their cross-section. This mask resulted in a well-defined triangular cross-section without mask redeposition and roughness along the waveguide edges as compared to the HSQ-Ti mask stack [Fig. \ref{fig:FigFab4-PMMANb}(a)(ii), (iii)]. This was due to the lower mask erosion rate of niobium as compared to HSQ during the oxygen RIBAE step, as well as the absence of the Ti layer that led to micromasking. Using a stage tilt of \(\alpha\) = 45\({}^{\circ}\) resulted in a top apex angle of \(\sim\) 53\({}^{\circ}\), with the discrepancy arising from mask erosion that narrowed the width of the mask.

To demonstrate the effectiveness of the improved triangular cross-section, we fabricated ring resonators with HSQ-Nb mask [Fig. \ref{fig:FigFab4-PMMANb}(b)] and compared them to those fabricated using HSQ-Ti mask. \cite{AtikianAPLPhot2017} After the RIBAE step, the diamond substrate with fabricated ring resonators was cleaned first with a mixture of nitric acid, phosphoric acid, and HF in 1:1:1 ratio to remove HSQ and Nb, followed by cleaning in piranha mixture, acetone, and methanol, before critical point drying with carbon dioxide. 

For ring resonators fabricated with HSQ-Nb mask stack, a total optical Q of 730 000 with an intrinsic optical Q of 776 000 were measured [Fig. \ref{fig:FigFab4-PMMANb}(c)], an improvement from the total optical Q’s of 286 000 measured from similar ring resonators fabricated with HSQ-Ti mask stack, \cite{AtikianAPLPhot2017} thus demonstrating that the improved smoothness of the triangular cross-section resulted in lower optical scattering losses.

\subsection{Nb mask via inversion of PMMA template (PMMA/Nb)}
However, metal masks possess grain structure, which subsequently transfers into diamond as sidewall roughness. [Fig. \ref{fig:FigFab4-PMMANb}(d)] To circumvent this, we devised an alternative patterning process through mask inversion [Fig. \ref{fig:FigFab4-PMMANb}(e)]. In this modified process, Poly(methyl methacrylate) (PMMA, 950 C4, MicroChem) was spun onto a cleaned diamond substrate multiple times to obtain a thickness of approximately 800 nm. The thickness of PMMA set the final thickness of the mask. The mask pattern was written in PMMA using EBL; due to the positive tone of the resist, this resulted in trenches and pillars where waveguides and holes were defined. After development with methyl iso-butyl ketone (MIBK) diluted in 1:3 ratio with isopropyl alcohol (IPA) for 60 seconds, the template was completed. The template was then inverted through sputtering of niobium such that the template is conformally covered, followed by an etchback to remove excess niobium on top of the PMMA template. 

To fully planarize the PMMA template through sputtering, the required material thickness is equal to half the width of the widest feature if the deposition is conformal. In our photonic crystal devices, this would correspond to the clamps of 3 \(\mu\)m width, which would require a 1.5 \(\mu\)m-thick layer of niobium to be sputtered. In the case of incomplete planarization, an etchback that uses conventional top-down etching would etch niobium at the bottom of unplanarized features, leading to an absence of mask in those regions. To avoid the need for a thick niobium layer, we employed an argon ion beam at grazing incidence for the etchback process instead. With the correct choice of ion beam incidence angle and thickness of the sputtered niobium layer, unplanarized features can be preserved by virtue of its sidewall coating shadowing the argon ion beam from reaching the bottom of the unplanarized region. The grazing-incidence argon ion beam milling was performed in the same Intlvac Nanoquest tool used for oxygen RIBAE, with the following parameters: argon flow 15 sccm, beam voltage 300 V, accelerator voltage 36 V, beam current 100 mA, ICP power 110 W, stage angle 20\({}^{\circ}\).

After the milling step to remove excess niobium on top of PMMA, thereby leaving the niobium mask surrounded by exposed PMMA, the PMMA was removed by immersion in Remover PG (MicroChem) for 24 hours to complete the mask definition process [Fig. \ref{fig:FigFab4-PMMANb}(f)(i)]. Device fabrication then proceeded with top-down oxygen etch followed by oxygen RIBAE with 200 V beam voltage. The niobium mask patterned in this way produced smooth sidewalls in the holes as inherited from the smoothness of the PMMA template, which then translated to smooth sidewalls in diamond as well after the anisotropic etch. However, after the oxygen angled RIBAE step, the holes became roughened. The roughening of holes was due to insufficient mask thickness to shadow the holes during the angled etching process [Fig. \ref{fig:FigFab4-PMMANb}(f)(ii)].

In turn, the insufficient mask thickness resulted from physical limitations in the sputtering process. This was because sputtering is not perfectly conformal; sharp convex corners at the top of trenches see a greater flux of incoming adatoms as compared to concave corners at bottom of trenches. Complete coating is possible for sufficiently low aspect ratios, given as the ratio of the height of a trench to its width. Above a critical aspect ratio, the disparity in coating at top corners of trenches compared to bottom corners would result in insufficient sidewall coating and voids within the mask. A mask with voids and incomplete sidewall coating would be compromised in its ability to withstand lateral erosion, especially over long durations of oxygen RIBAE for telecommunication wavelength photonic crystals.

The optical reflection spectrum of a telecommunication photonic crystal, fabricated with template-patterned Nb mask, is shown in Fig. \ref{fig:FigFab4-PMMANb}(g). The telecommunication wavelength photonic crystals fabricated with the modified niobium masking process and RIBAE were measured to have lower intrinsic optical Q of 39 700, compared to 270 000 measured in telecommunication photonic crystals fabricated with HSQ-Ti mask and Faraday cage angled etching. \cite{BurekOptica2016} Here, intrinsic optical Q was compared instead of total optical Q to exclude variations in waveguide coupling rates, as the two devices were measured via different means (reflection vs transmission respectively). The degraded Q’s can be attributed to roughened holes arising from insufficient Nb mask thickness and hole shadowing during oxygen RIBAE, even though the Nb mask gave a better cross-section compared to HSQ-Ti mask in Faraday cage angled etching.

\subsection{HSQ-alumina mask} \label{chap:ibe-subsec:HSQ-alumina}
\begin{figure}
\centering\includegraphics[width=\textwidth,keepaspectratio]{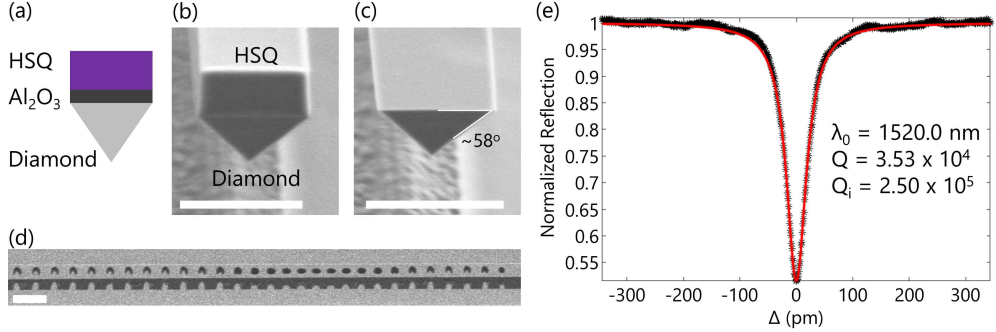}
\caption[RIBAE of photonic crystal with telecommunication resonance wavelength using HSQ/alumina mask]{RIBAE of photonic crystal with telecommunication resonance wavelength using HSQ/alumina mask. (a) Schematic of HSQ-alumina (Al\({}_{2}\)O\({}_{3}\)) mask stack for angled etching. Here the alumina layer is 1 nm thick and serves as an adhesion layer. (b) Scanning electron microscope (SEM) image showing cross-section of completely etched cantilever, prior to HSQ mask removal. (c) SEM image of the same cantilever in (b), after HSQ mask removal, with a final etch angle close to 58\({}^{\circ}\). (d) SEM image of photonic crystal after angled etching with HSQ-alumina mask. (e) The optical reflection spectrum of photonic crystal fabricated with HSQ-alumina mask, with optical resonance at 1520.0 nm, total optical Q of 35 300, and intrinsic optical Q of 250 000. Scale bars in (b) – (d) represent 1 \(\mu\)m. All SEM images were taken at 60\({}^{\circ}\) tilt.
\label{fig:FigFab5-HSQAlumina}}
\end{figure}

The low measured optical Q’s of telecommunication wavelength photonic crystals using PMMA/Nb mask as compared to those achieved with HSQ-Ti mask (with Faraday cage angled etching) shows that a thick mask is needed to shadow the photonic crystal holes from being etched and produce high optical Q’s. However, it is difficult to deposit thick layers of commonly used hard mask materials without accumulating stress and/or without grain structure. A compromise was reached by using a thick HSQ layer, but with a thin alumina underlayer instead of a metallic underlayer as in the HSQ-Ti process [Fig. \ref{fig:FigFab5-HSQAlumina}(a)]. An alumina layer approximately 1 nm thick was deposited via atomic layer deposition (Cambridge NanoTech) after the diamond substrate cleaning step, but prior to spin coating of 1 \(\mu\)m of HSQ. After EBL and development, a short Ar/Cl\({}_{2}\) etch was used to remove the alumina layer prior to top-down diamond etching.

We first fabricated cantilevers with the HSQ-alumina mask stack to demonstrate smooth triangular cross-sections during and after the RIBAE process [Fig. \ref{fig:FigFab5-HSQAlumina}(b), (c)], resembling that from the single-stack PMMA/Nb process described earlier [Fig. \ref{fig:FigFab4-PMMANb}(a)(ii)]. We used using the recipe with a lower beam voltage of 150 V to achieve a lower erosion rate of HSQ. The corresponding lateral etch rate of diamond with this recipe is approximately 2.3 nm/min. The amorphous nature of the alumina underlayer means that there is no grain structure that results in diamond sidewall roughness via top-down diamond etching, while the thinness of the layer eliminates mask redeposition. Therefore, the combined HSQ-alumina mask stack can be considered an effective single-stack mask [Fig. \ref{fig:FigFab5-HSQAlumina}(b)]. 

The cross-section of a cantilever fabricated using the HSQ-alumina mask has a top apex angle of \(\sim\) 58\({}^{\circ}\), [Fig. \ref{fig:FigFab5-HSQAlumina}(c)] which is 13\({}^{\circ}\) off of the stage tilt of \(\alpha\) = 45\({}^{\circ}\) used. The greater discrepancy for the HSQ-alumina mask using RIBAE at 150 V beam voltage, compared to 8\({}^{\circ}\) for HSQ-Nb mask and 200 V beam voltage, signifies that lateral mask erosion is much more pronounced without a metal underlayer. This is in spite of a lower beam voltage recipe used with the intention of reducing physical sputtering of HSQ by oxygen ions. However, the vertical mask erosion is reduced at lower beam voltages, as shown by the successful etch of telecommunication wavelength OMCs using HSQ-alumina mask, compared to unsuccessful etch using HSQ-Ti mask.

Telecommunication-wavelength photonic crystals were then fabricated with HSQ-alumina mask with the process outlined above, using the same design that was used for the PMMA/Nb process, but with the waveguides widened to account for mask erosion during RIBAE. The waveguides were deliberately made wider at the lithography step to account for lateral erosion during RIBAE, given the significant HSQ erosion rate observed when HSQ-based masks were used. A scanning electron micrograph (SEM) of a completed device is shown in Fig. \ref{fig:FigFab5-HSQAlumina}(d). For a photonic crystal fabricated with the HSQ-alumina mask, optical reflection spectroscopy revealed an intrinsic optical Q of 250 000. This is nearly an order of magnitude improvement from the Q of 39 700 measured from photonic crystals fabricated with PMMA/Nb mask, showing that a sufficiently thick mask is needed for high optical Q’s.

\section{Conclusion}
The lack of high-quality diamond materials grown through heteroepitaxial methods has motivated the development of nanofabrication techniques in bulk substrates. The angled-etching technique is studied in particular here owing to the success of fabricating quantum photonic devices in the visible wavelength range using the reactive ion beam angled etching (RIBAE) technique. To extend this success of angled-etching to devices in the telecommunication wavelength range, various mask choices were evaluated based on their ability to produce smooth triangular cross-sections reliably. The cross-sections should have no roughness due to mask redeposition or insufficient shadowing during the angled etching process. 

This work outlines the requirements for a good mask stack: it should be sufficiently thick, have high selectivity, good aspect ratio tolerance, high resolution and smooth sidewalls. However, we found that it was not possible to satisfy all five criteria simultaneously through measurements of optical Q’s of photonic devices fabricated using various masks. We summarize the measured optical Q's in Table \ref{tab:table1}. Masks based on HSQ (HSQ-Ti, HSQ-Nb, HSQ-alumina) can be made sufficiently thick with good resolution and aspect ratio tolerance, but suffers from low selectivity due to its substantial erosion rate in RIBAE. Additionally, the HSQ-Ti masks showed redeposition of mask, while the Nb layer in HSQ-Nb masks exhibited grain boundaries, leading to sidewall roughness on the outer edges of devices and/or inner walls of holes. In the Nb mask made from PMMA template inversion, even though the PMMA provided high resolution and ensured smooth sidewalls of the templated Nb mask, which in turn provided high selectivity, the sputtering process limits the highest aspect ratio and mask thickness possible. Finally, even though the HSQ-alumina mask had low selectivity, it gave smooth sidewalls by virtue of the amorphous alumina layer. Despite not satisfying all five criteria, telecommunication wavelength photonic crystals fabricated with RIBAE using HSQ-alumina mask gave the highest intrinsic optical Q’s compared to the other mask choices, and this Q was comparable to those previously demonstrated with Faraday cage angled etching. Therefore, we conclude that with the right choice of mask, RIBAE can produce the same quality devices as Faraday cage approach, with added advantages of producing these devices over areas that are orders of magnitude larger, and with much better uniformity and yield. 

\begin{table}
\centering
\caption{\label{tab:table1}Table comparing optical Q-factors between different devices fabricated in diamond via angled etching, using different masks.}
\begin{adjustbox}{width=1\textwidth}
\begin{tabular}{ccccc}
\hline
Device type         & Etching method & Mask material(s) & Wavelength (nm) & Optical Q \\
\hline
Racetrack resonator & Faraday cage   & HSQ-Ti           & 1648.7     & 151 000 (total, \cite{BurekNComm2014})     \\
Racetrack resonator & RIBAE          & HSQ-Ti           & $\sim$1600 & 286 000 (total, \cite{AtikianAPLPhot2017})     \\
Racetrack resonator & RIBAE          & HSQ-Nb           & 1621.0     & 730 000 (total, this work) \\
Photonic crystal    & Faraday cage   & HSQ-Ti           & 1529.2     & 270 000 (intrinsic, \cite{BurekOptica2016}) \\
Photonic crystal    & RIBAE          & HSQ-Ti           & nil        & nil\\
Photonic crystal    & RIBAE          & PMMA/Nb          & 1566.2     & 39 700 (intrinsic, this work)  \\
Photonic crystal    & RIBAE          & HSQ-alumina      & 1520.0     & 250 000 (intrinsic, this work) \\
\hline
\end{tabular}
\end{adjustbox}
\end{table}

While diamond is specifically studied for its relevance to quantum photonics, the same considerations for choice of mask can also be applied to bulk materials where no heteroepitaxial or thin-film-on-insulator platforms are available. However, limitations on mask thicknesses prevent devices larger than telecommunication wavelength scale to be fabricated. Devices wider than those intended for telecommunication wavelengths would take a longer RIBAE etch to be fully suspended, which in turn require a thicker mask to account for lateral and top-down mask erosion. However, thick deposited materials (> 1 \(\mu\)m) typically exhibit large stresses, which can lead to delamination from substrates. In addition, pattern transfer into a thick layer requires a first mask layer with sufficient selectivity and high resolution, especially for devices with small features such as photonic crystals. These considerations would preclude common masking methods involving pattern transfer or lift-off, and would require fine-tuning of deposition processes to minimize stresses, or innovative deposition techniques that can provide a thick mask material with low accumulated stress, high resolution, and large aspect ratios.

\begin{backmatter}
\bmsection{Funding}
National Science Foundation (CQIS ECCS-1810233, EFRI ACQUIRE 5710004174, NNCI ECCS-2025158, STC DMR-1231319); Office of Naval Research (MURI N00014-15-1-2761, MURI N00014-20-1-2425); U.S. Department of Energy (HEADS-QON DE-SC0020376).

\bmsection{Acknowledgments}
C.C. contributed to this work prior to joining IMRE. B.M. contributed to this work prior to
joining AWS. We wish to thank H. A. Atikian and M. J. Burek for stimulating discussions, and E. Cornell for help with device fabrication and characterization. This work was supported by NSF (EFRI ACQUIRE Grant No. 5710004174, CQIS grant No. ECCS-1810233, STC grant No. DMR-1231319), ONR (MURI on Quantum Optomechanics, Grant No. N00014-15-1-2761 and N00014-20-1-2425) and DOE (HEADS-QON Grant DE-SC0020376). C. C. acknowledges support from the National Science Scholarship of the Agency for Science, Technology and Research (A*STAR), Singapore. This work was performed in part at the Harvard University Center for Nanoscale Systems (CNS); a member of the National Nanotechnology Coordinated Infrastructure Network (NNCI), which is supported by the National Science Foundation under NSF award no. ECCS-2025158. CNS is part of Harvard University.

\bmsection{Disclosures}
The authors declare no conflicts of interest.

\bmsection{Data availability} Data underlying the results presented in this paper may be obtained from the authors upon reasonable request.

\end{backmatter}


\bibliography{main}

\end{document}